\documentclass[12pt]{iopart}
\usepackage{graphicx}
\usepackage{pstricks,pst-node}
\usepackage{pst-node,pst-text,pst-3d} % Used by prosper etc.
\usepackage{verbatim}
\input epsf.tex 
\def\ii{{\'{\i}}}

\def\C{{\af C}}

\def\CE{{\cal E}}
\def\CO{{\cal O}}

\def\CN{{\cal N}}\def\CU{{\cal U}}
\def\CU{{\cal U}}

\def\be{\begin{eqnarray}}    
\def\ee{\end{eqnarray}}
\def\Dsl{\,\raise.05ex\hbox{/}\mkern-9.5mu D}
\def\boxeqn#1{\vcenter{\vbox{  \hrule height2pt \hbox{\vrule
width 2pt \kern3pt\vbox{\kern3pt
\hbox{${\displaystyle #1}$}\kern3pt}\kern3pt\vrule width 2pt}\hrule height2pt}}}
\def\back{{{\raise.4em\hbox{$\, _\backslash\,$}}}}

 2

\font\blackboard=msbm10 \font\blackboards=msbm7
\font\blackboardss=msbm5
\newfam\black
\textfont\black=\blackboard
\scriptfont\black=\blackboards
\scriptscriptfont\black=\blackboardss

\def\frac#1#2{{#1\over #2}}

\def\big R{{\hbox{{\bigfield R}}}}
\def\bbig R{{\hbox{{\bbigfield R}}}}

\font\afm=msbm12

\def\C{\hbox{\afm C}}

\def\Rm{\hbox{\afm R}}
\def\I{\hbox{\afm I}}

\mathchardef\imath="717B
\def\inbar{\,\vrule height1.5ex width.4pt depth0pt}
\def\IB{\relax{\rm I\kern-.18em B}}
\def\IC{\relax\hbox{$\inbar\kern-.3em{\rm C}$}}
\def\ID{\relax{\rm I\kern-.18em D}}
\def\IE{\relax{\rm I\kern-.18em E}}
\def\IF{\relax{\rm I\kern-.18em F}}
\def\IG{\relax\hbox{$\inbar\kern-.3em{\rm G}$}}
\def\IH{\relax{\rm I\kern-.18em H}}
\def\II{\relax{\rm I\kern-.18em I}}
\def\IK{\relax{\rm I\kern-.18em K}}
\def\IL{\relax{\rm I\kern-.18em L}}
\def\IM{\relax{\rm I\kern-.18em M}}
\def\IN{\relax{\rm I\kern-.18em N}}
\def\IO{\relax\hbox{$\inbar\kern-.3em{\rm O}$}}
\def\IP{\relax{\rm I\kern-.18em P}}
\def\IQ{\relax\hbox{$\inbar\kern-.3em{\rm Q}$}}
\def\IR{\relax{\rm I\kern-.18em R}}
\font\cmss=cmss10 \font\cmsss=cmss10 at 10truept%!!! should be 7pt
\def\IZ{\relax\ifmmode\mathchoice
{\hbox{\cmss Z\kern-.4em Z}}{\hbox{\cmss Z\kern-.4em Z}}
{\lower.9pt\hbox{\cmsss Z\kern-.36em Z}}
{\lower1.2pt\hbox{\cmsss Z\kern-.36em Z}}\else{\cmss Z\kern-.4em Z}\fi}
\def\IGa{\relax\hbox{${\rm I}\kern-.18em\Gamma$}}
\def\IPi{\relax\hbox{${\rm I}\kern-.18em\Pi$}}
\def\ITh{\relax\hbox{$\inbar\kern-.3em\Theta$}}
\def\IOm{\relax\hbox{$\inbar\kern-3.00pt\Omega$}}

\def\CO{{\cal O}}

\def\CH{{\cal H}}

\def\CE{{\cal E}}

\def\CU{{\cal U}}

\def\CN{{\cal N}}
\def\CD{{\cal D}}

\begin{document}

\title{Vacuum Boundary Effects}

\author{M. Asorey and  J. M.  Mu\~noz-Casta\~neda}

\address{Departamento de F\'\i sica Te\'orica. Facultad de Ciencias.
Universidad de Zaragoza, 50009 Zaragoza. Spain}
\ead{asorey@unizar.es}
\begin{abstract}

The  effect of  boundary conditions on the  vacuum  structure of quantum
field theories is analysed from a  quantum information  viewpoint.
In particular, we analyse the role of boundary conditions on    boundary
entropy and entanglement entropy.  The analysis of boundary effects
 on massless free field theories points out   the  relevance of boundary conditions 
as a new rich source of  information about the vacuum  structure. In all cases  the entropy does 
not increase along the  flow  from the ultraviolet to the infrared.

\end{abstract}


%{\it Keywords:}{Vacuum structure, boundary conditions, conformal invariance, boundary entropy}
\maketitle

\section{Introduction}

In quantum field theory the vacuum state encodes all physical properties of the  theory.
Indeed, any other state can be generated by the action of field operators on the vacuum.
In particular, the effects generated by  non-trivial  topological  structures of space or 
change of boundary conditions can be directly analysed from  the
 changes induced on the vacuum structure. Among the most famous
 vacuum effects  are the phenomenon of  spontaneous symmetry breaking 
and the Casimir effect \cite{casimir}.

In particle physics, the main interest  usually focuses  on the   behaviour of Green's 
and other quantum field correlation functions at short distances  which provides 
information about  high energy particle scattering processes. These observables are  very insensitive
to space topology or field boundary conditions \cite{karpacz}. However, for strongly correlated 
or confining theories  long distance properties become very important, for instance, 
to point out the existence or not of confinement or mass gap. The existence of deconfining 
transitions in those theories  (e.g. non-abelian gauge theories) can  be directly extracted 
from the analysis of the structure of the vacuum state. Another rich source of information about
the theory is encoded in the behaviour of  non-local observables like  free energy
or entropy that can be defined by exploiting  analogies with thermodynamics.

The interest on observables of this type   has been recently boosted by the development of
quantum information theory. The entanglement entropy \cite{sorkin} provides a good  measure 
of the vacuum entanglement structure. It can also be used to point out the existence of phase 
transitions since it is  unbounded  for critical systems and
bounded for systems with a finite mass gap \cite{guifre}. It has been also pointed out that the 
confinement mechanism might be related to vacuum entanglement \cite{kleb}.
Another thermodynamic observable, the boundary entropy
\cite{cardy}\cite{affleck} is related to the number of boundary states. Both  new types of  entropy
do not scale with the volume of the space, unlike the standard bulk entropy  and other  extensive quantities.
The entanglement entropy scales in the critical case with the area of the boundary
where the fluctuating modes of the vacuum are traced out \cite{sorkin}\cite{srednicki}. This 
behaviour is characteristic of black hole physics and is one of the key features
 of the AdS/CFT correspondence.

By their own nature it is quite possible that both new entropies shall depend on the global properties
of the configuration space.  In this note we analyse  the dependence of those quantities 
on the space topology and field boundary conditions as well as its physical implications for 
quantum field theories.

\section{Boundary conditions and conformal invariance}

Let us consider   a   real  scalar free field theory defined in a bounded domain $\Omega$ in $\IR^D$
 with regular and smooth boundary $\partial \Omega$.
The quantum dynamics is governed by the Hamiltonian
\be
\CH=-\frac12\left\|\frac{\delta}{\delta \phi}\right\|^2 +\frac12\left(\phi,\sqrt{-\Delta+m^2}\, \phi\right).
\label{prima}
\ee
Unitarity requires that $\CH$ has to be selfadjoint. In particular, this implies that
one must fix the   boundary conditions of the fields $\phi$ in a way that 
the Laplace-Beltrami operator $-\Delta$ is selfadjoint and positive.
The boundary conditions  which define a selfadjoint operator $-\Delta$ 
are given by \cite{aim}
\be
{\phantom{\Bigl[}  
\varphi- i\, \dot\varphi ={   U} \left(\varphi+ i\, \dot\varphi
\right)\phantom{\Bigr[} }
\label{const}
\ee
in terms of an  unitary operator $U\in \CU(L^2(\partial\Omega,\C))$
which acts on the boundary values $\varphi$ of the quantum fields $\phi$ and 
their  normal derivatives $\partial_n\varphi=\dot\varphi$.
Notice that not  all unitary operators give rise to  positive  Laplace-Beltrami operators, but
to have a consistent quantum field theory for all values of $m$ one needs to consider only boundary
conditions which satisfy both requirements. The set of boundary conditions which are compatible with unitarity is 
given by unitary matrices $U$ with eigenvalues $\lambda={\rm e}^{i \alpha}$
in the upper unit semi-circumference  $0\leq \alpha\leq \pi $.
For  a single real scalar field  defined on the two-dimensional space-time $\IR\times[0,L]$
  the set of compatible boundary conditions is a four-dimensional manifold  which can be 
covered by two charts parametrised by
\be
L \begin{pmatrix} {\dot{\varphi}(0) \cr \dot{\varphi}(L)}
\end{pmatrix}=
A \begin{pmatrix} {\varphi(0) \cr \varphi(L)} 
\end{pmatrix}
\label{uno}
\ee
where $A=-i(\I-U)/(\I+U)$ is any  hermitian matrix with $A\geq 0$ , and 
\be
 \begin{pmatrix} { \varphi(L) \cr L \dot\varphi(L)}
\end{pmatrix}=
B \begin{pmatrix} {\varphi(0) \cr L \dot\varphi(0) }
\end{pmatrix}
\label{dos}
\ee
where $B=\begin{pmatrix} {a & b \cr c & d}\end{pmatrix}$ is any  real matrix   
with $ad+bc=-1$, $ ac \leq 0$ and $bd \leq 0$.

In the  massless case $m=0$ the theory is conformally invariant.
However, most of the compatible  boundary conditions (\ref{uno}) (\ref{dos}) 
break conformal invariance \cite{cardy}. Only the boundary conditions corresponding to unitary 
matrices $U$ with eigenvalues $\pm 1$ preserve conformal invariance \cite{agm, agm2}.
In the two-dimensional case the set of conformally invariant boundary conditions
\be 
\left\{\I,-\I, 
U_\alpha= \begin{pmatrix} {\cos \alpha &\sin \alpha  \cr \sin \alpha&-\cos \alpha }
\end{pmatrix}; \alpha \in (0,2\pi] \right\}\subset U(2),
\label{uni}
\ee
is given by Neumann ($U=\I$), Dirichlet ($U=-\I$) and quasiperiodic ($U_\alpha$)
boundary conditions \cite{agm}.
 All other compatible boundary conditions break conformal invariance and are not invariant under  
renormalization group transformations. They  describe renormalised trajectories of  the renormalization group 
flowing towards one of the conformally invariant boundary conditions \cite{agm2}. 
% This boundary renormalization group flow might also have dynamical meaning as the
% adiabatic evolution under the one-parametric family of Hamiltonians connected by 
% scale transformations.

\section{Boundary effects in  conformal field theories}

The infrared properties of quantum field theory are very sensitive to quantum field boundary
conditions \cite{karpacz}. In particular, the physical properties of the quantum 
vacuum, free energy  and  vacuum energy exhibit  a very strong dependence on 
the type of boundary conditions. 

The vacuum state of the free field theory is gaussian
\be
\Psi(\phi)= \CN \ {\rm e}^{\displaystyle -\frac12 (\phi, \sqrt{-\Delta+m^2}\, \phi)}
\ee 
and the vacuum energy density 
$\CE_0= \tr \sqrt{-\Delta+m^2}\,$
is ultraviolet divergent. However, for finite cylindric domains of the form $S^{D-1}\times[0,L]$
 the finite size corrections $\epsilon_c$ of the  asymptotic expansion of  the vacuum energy density  
for  large values of cylinder  base radius   $\Lambda$ and generatrix $L$  with $\Lambda>>L>>1$
\be
\CE_0=\epsilon_B+\epsilon_b \frac1{L} 
+ \frac1{L^{D+1}} {\epsilon_c(m L)} +\CO\left(\frac1{\Lambda}\right)
\label{ve} 
\ee
are  not divergent  \cite{casimir}.
In the massless limit $m\to 0$ the coefficient  $\epsilon_c$ of this term
becomes universal (i.e. independent of $L$) but  is highly dependent on the boundary 
conditions\footnote[7]{The absence of logarithmic corrections 
$\CO(\log L)$ is due to the topology of the boundary.
In general those  corrections   spoil the universal character of 
the $\CO(1)$ term \cite{peschel} }.
For instance, in two dimensions  for 
 quasi-periodic boundary conditions this first finite size correction is  
\be
 \epsilon_c=
\frac{\pi}{12}- {\pi} \left[ \frac{\alpha}{2\pi}-\frac{3}{4} 
\right]^2.
\label{qp}
\ee
The values and signs of this finite size contribution to the energy are very different
for periodic ($\alpha=\pi/2,\epsilon_c=-\pi/6$), antiperiodic ($\alpha=3\pi/2,\epsilon_c=\pi/12$) and Zaremba  
($\alpha=\pi ,\epsilon_c=\pi/48$) boundary conditions \cite{bfsv}-\cite{klp}. In higher dimensions we 
have  for  domains of the form  $S^{1}\times[0,L]$  the  values of $\epsilon_c$:   
 $-\zeta(3)/(2\pi)$) for periodic,  
$3\,\zeta(3)/(8\pi)$ for antiperiodic and  
$3\, \zeta(3)/( 6 4\pi)$ for Zaremba boundary conditions, where $\zeta(3)=1.2020569$ is 
 Ap\'ery's constant \cite{elizalde}. Similarly, 
in three-dimensional cylindric domains   $S^{2}\times[0,L]$ we have for the same boundary conditions
$-\pi^2/90$,  $7 \pi^2/720$ and  
${7 \pi^2/11520}$, respectively \cite{elizalde}.

In a similar manner the free energy  of the system at finite temperature $1/T$ 
with the boundary conditions (\ref{const})  has 
the following asymptotic expansion for large volumes and low temperature $0<<L<<T<<\Lambda$ \cite{affleck, Cardy},
\be
 f=-\frac{  \hbox{log Z}}{\Lambda^{D-1} L}={ f_B}\, T   + f_b\frac{T}{L} \,   + \frac{T}{L^{D+1}} {f_c(m L)} 
+ \CO\left(\frac1{T}, \frac1{\Lambda}\right),
\ee
where $f_B=\epsilon_B$, $f_b=\epsilon_b$ and $f_c=\epsilon_c$. This is in agreement with the 
asymptotic expansion of vacuum energy density (\ref{ve}) and for the same reason does not present 
any logarithmic dependence in the  smaller transverse size scale  $L$.

In the  asymptotic regime of  low temperature and large volumes  $0<<T<<L<<\Lambda$ we have
\be
 f=-\frac{  \hbox{log Z}}{\Lambda^{D-1} L} ={ f_B}\,  T   +  \, \frac{1}{T^{D}} \tilde{f}_c(mT)
 + \CO\left(\frac1{L}, \frac1{\Lambda}\right).
\ee

There is a similar expansion for the entropy
$$
\hbox{S}= (1-T\partial_T)\log \hbox{Z}= - (D+1)\frac{\hbox{ }\Lambda^{D-1} L} {T^D}\tilde{f}_c(mT)% - \frac{\hbox{ }R^{D-1} L} {T^D}\tilde{f}_c(mT)
+\frac{\hbox{ }m \Lambda^{D-1} L} {T^{D-1}}\tilde{f}'_c(mT) +% { s_b} +
\, { s_b }+  \CO\left(\frac1{L}, \frac1{\Lambda}\right).
$$
The third   term  of this expansion $s_b$, known as boundary entropy \cite{cardy}\cite{affleck},
is finite  and depends on the boundary conditions of the fields. In two dimensional conformal theories
this entropy $s_b=\log g$ can be formally associated with the number  of boundary states
$g$ \cite{cardy},  but in many cases $g={\rm e}^{\log s_b}$ is not integer  and does not
correspond to a simple counting of boundary states \cite{affleck}. It has been conjectured
that  the quantities  $g$ and $s$ evolve with the renormalization group flow 
 in a non-increasing way \cite{affleck} 
$$s_{ _{UV}}\geq s_{ _{IR}},\quad g_{ _{UV}}\geq g_{_{IR}} $$
as it corresponds to any  type of thermodynamic entropy \cite{affleck}\cite{friedan}. 
This conjecture is known as $g$-theorem and has been verified in many cases 
\cite{affleck2}\cite{friedan} although not yet proved for the boundary renormalization group flow.

The conjecture can be verified in the case of a two-dimensional free real scalar field defined on $\Rm\times [0,L]$. 
The partition function for anti-periodic boundary conditions, once properly renormalised, can be exactly calculated and 
it is given by
\be
 Z_a= q^{\frac1{24}}\prod_{n=1}^\infty (1-q^{n-\frac1{2}})^{-2}= 
 \frac12\,\tilde{q}^{-\frac1{12}}\prod_{n=1}^\infty (1-\tilde{q}^{2n-1})^{2},
\label{antiperiodic}
\ee
where  ${q}=e^{-2\pi T/L}$ and $\tilde{q}=e^{-2\pi L/T}$.
From (\ref{antiperiodic}) it follows that Casimir coefficient is  in this case $\epsilon_c=\frac{\pi}{12}$. For 
Zaremba boundary conditions \cite{klebanov} we have
\be
 Z_z= q^{\frac1{96}}\prod_{n=1}^\infty (1-q^{\frac{n}{2}-\frac1{4}})^{-1}
 = \hbox{$\frac1{\sqrt{2}}$}\,\tilde{q}^{-\frac1{12}}\prod_{n=1}^\infty (1-\tilde{q}^{4{n}-2}),
\label{zaremba}
\ee
which leads to the Casimir coefficient  $\epsilon_c=\frac{\pi}{48}$. 

For periodic boundary conditions there are zero modes which generate infrared divergences. The
partition function
(density)   is given by \cite{polchinski}
\be 
z_{p}= \hbox{$\sqrt{\frac{L}{2\pi T}}$}\, 
{q}^{-\frac1{12}}\prod_{n=1}^\infty (1-{q}^n)^{-2}=\hbox{$ \sqrt{\frac{T}{2\pi L}}$}\, 
\tilde{q}^{-\frac1{12}}\prod_{n=1}^\infty (1-\tilde{q}^n)^{-2}.
\label{peri}
\ee

But, the infrared problem  is so severe that affects  the consistency of the  theory  \cite{Coleman}.
In any quantum field theory the Schwinger functions  must satisfy the   Osterwalder-Schrader reflection positivity
property in order to preserve unitarity and causality. However, in a  free theory of two-dimensional massless  bosons 
the two point function is neither positive nor reflection positive \cite{gift}.  One way of solving all these problems is to consider
a compactification of the scalar field $\Phi= {\rm e}^{i \phi/R}$ to a circle of unit radius. In that case the correlators
of the compactified field $\Phi$ satisfy the reflection positivity requirement and theory becomes consistent \cite{gift}.

In that case the partition function  acquires some additional contributions dues
the compactification of zero-modes. 
In particular,  these contributions  give rise to the following partition function 
\be 
Z_{p}^R &=&  {q}^{-\frac1{12}}\prod_{n=1}^\infty (1-{q}^n)^{-2} 
\sum_{n,m=-\infty}^\infty {q}^{ \pi R^2   n^2+ \frac{m^2}{4 \pi R^2 } } 
\label{peri2}\\  
&=& \tilde{q}^{-\frac1{12}}\prod_{n=1}^\infty (1-\tilde{q}^n)^{-2} %Z_p
\sum_{n,m=-\infty}^\infty \tilde{q}^{ \pi R^2   n^2+ \frac{m^2}{4 \pi R^2 } }
\label{peri22}
\ee
for periodic boundary conditions.

However, for the rest of quasiperiodic boundary conditions ($\alpha\neq \pi/2 $) there is no 
contribution of the compactification of zero modes and the partition function is directly given by
\be 
 Z_{a}^R &=&
 q^{\frac1{24}-\frac12({\epsilon-\frac12})^{2}}  \prod_{n=-\infty}^\infty \left(1-q^{|n-\epsilon|}\right)^{-1}\\
&=&
\hbox{$ $}\tilde{q}^{-\frac1{12}}\ \left({2\, {\sin}{\pi \epsilon} }\right)^{-1}
\prod_{n=1}^\infty \left|1-e^{2\pi \epsilon i}\tilde{q}^{n}\right|^{-2}
\label{peri5}
\ee
where $\epsilon=|\frac{\alpha}{2\pi }- \frac14|$. 
In particular, this means that  for antiperiodic and Zaremba boundary conditions  there is no modification 
of (\ref{antiperiodic}) and  (\ref{zaremba}), respectively.

 For Neumann  boundary conditions the partition function is also modified by the
presence of compact zero modes
\be 
Z_{N}^R &=& \,{q}^{-\frac1{48}} \prod_{n=1}^\infty (1-{q}^{{n/2}})^{-1} 
\sum_{n=1}^\infty {q}^{ \frac{n^2}{4 \pi R^2}}\\
&=& \sqrt{\pi} R
\,  \tilde{q}^{-\frac1{12}} \prod_{n=1}^\infty (1-\tilde{q}^{{2n}})^{-1} 
\sum_{n=1}^\infty \tilde{q}^{ \pi R^2  n^2} %n^2 | n-2\pi m T/L|^2}{{ T}}
\label{neumann2}
\ee
in a similar way  that for the theory with Dirichlet  boundary conditions, where 
\be 
Z_{D}^R&=& \,  {q}^{-\frac1{48}}\prod_{n=1}^\infty (1-{q}^{{n/2}})^{-1} \!\! 
\sum_{m=-\infty}^\infty {q}^{ \pi R^2 m^2}{ }\\
&=& \, \hbox{$ \frac{1}{2R\sqrt{\pi}}$}\,   \tilde{q}^{-\frac1{12}}\prod_{n=1}^\infty (1-\tilde{q}^{{2n}})^{-1} \!\! 
\sum_{m=-\infty}^\infty \tilde{q}^\frac{m^2}{4 \pi R^2 }. 
\label{dirichlet2}
\ee

 The boundary entropy can easily be computed for all those cases and the results are:

\begin{equation}
\begin{array}{llll}
s^\alpha_{b} &=-\log{(2\sin \pi\epsilon}) \  &g_\alpha=(2 \sin \pi \epsilon)^{-1} & \hbox{ quasiperiodic b.c.} \cr
s^D_{b}&=-\log 2R\sqrt{\pi}
\ & g_{_D}=({2R\sqrt{\pi}})^{-1} & \hbox{  Dirichlet b.c.}\\ s^Z_{b}&=-\frac12\log 2
\  & g_{_Z}={2}^{-\frac1{2}} & \hbox{ Zaremba b.c.} \\
s^N_{b}&=\phantom{-}\log R\sqrt{\pi}
\ & g_{_N}={R\sqrt{\pi}}{} &\hbox{  Neumann b.c.}
\end{array}
\end{equation}

The singularity observed for quasiperiodic boundary conditions at $\epsilon=0$ is due to 
the existence of zero-modes which once properly incorporated into the compact theory
give rise to the correct value for periodic boundary conditions (\ref{peri2}) (\ref{peri22}) with vanishing boundary entropy.
Notice also that  $g_{_Z}= \sqrt{g_{_D} g_{_N}}$ as corresponds to the factorisation property of counting boundary states.

The g-theorem holds along the renormalised flow of Robin boundary conditions 
$$
 U=\begin{pmatrix} {{\rm e}^{i \beta_0} &0  \cr 0&{\rm e}^{i \beta_L}  }
\end{pmatrix},$$
which interpolate 
between Dirichlet ($U=-\I$) to Neumann  ($U=\I$) boundary conditions through  Zaremba  ($U=\sigma_3$)
 boundary conditions \cite{agm3}
$$ g_{_D} > g_{_Z}> g_{_N}$$
provided that  $R< 1/\sqrt{2\pi}$. The  boundary entropy exhibits 
a monotone behaviour similar to that  of   the central charge or the bulk entropy.

\section{Entanglement Entropy}

There is another type of   entropy associated to the vacuum state of
a field theory. If  we ignore some  field degrees of freedom of the theory 
 one can consider the effective physical (mixed) states by tracing out those
 degrees of freedom. In this way  mixed states with finite
entropies can effectively appear in quantum field theory at zero temperature 
from pure states. The mechanism of tracing out degrees of freedom is a kind of  quantum version of the
renormalization group.  In particular, the vacuum state generates by this mechanism a family of mixed states 
whose entropies provide measures of its degree of entanglement. These
 mixed states are  generated by  integration of
the fluctuating modes of the vacuum state $\Psi_0$  in  bounded domains $\Omega_1$ of the physical space
$\Rm^D$ \cite{sorkin},  i.e.
\be
\rho_{_{\Omega_1}}=\int_{\Omega_1}  \Psi_0^\ast \Psi_0(x) d^D x.
\ee
The entropy  of this  state  $ S_{\Omega_1}= -Tr \,\rho_{_{\Omega_1}} \log \rho_{_{\Omega_1}}$
 (vacuum entanglement entropy) is  ultraviolet 
divergent, but once regularised  exhibit a very interesting asymptotic
behaviour which is similar to that of the boundary entropy analysed in the previous section 
\cite{srednicki}\cite{friedan}\cite{callan}\cite{dowker}\cite{kabat}. For massless scalar
theories  the entropy presents the following asymptotic 
behaviour
\be
\quad  S_{\Omega_1}=  
\sum_{i=0}^{D-1} C_i\,  \left(\frac{L_1} {a}\right)^{i}+\CO\left(\frac{a}{L_1} \right),
\label{al}
\ee
in terms of the diameter $L_1$ of $\Omega_1$ and
the  ultraviolet short distances cut-off  $a$  introduced to split apart  the domain $\Omega_1$ and its complement 
$\Rm^D \hbox{\, $\backslash$ \,}  \Omega_1$.  In the three-dimensional case, this asymptotic behaviour follows
 an area law similar to the black hole area law \cite{sorkin,srednicki}.
In general, for $D>1$ the coefficients $C_i$ are not universal because
 they are regularization dependent. However,
for one-dimensional spaces, although the formula (\ref{al})
suggests that $C_0$ could be universal, it does not happen.  In fact, the asymptotic behaviour 
of the entanglement entropy is not given in that case by
(\ref{al}) because that entropy   acquires a leading logarithmic correction
\be
 S_{\Omega_1}=C \log\frac{L_1} {a} +C_0,
\label{eone}
\ee
which obviously implies that  the constant term is highly dependent on the regularization method.
However, it turns out that the value of the coefficient of this logarithmic term $C$ is  universal and equal
to $1/3$ of the central charge $c$ of the conformal invariant theory. In the case  of a massless scalar
boson $c=1$ and 
$C=1/3$ \cite{holz}. The question is whether
this value is dependent or not  on the boundary conditions of the fields  when 
the  theory  is defined on a large bounded domain $\Omega\supset\Omega_1.$
It is remarkable that  coefficient $c_1=1/3$ turns out to be  independent of the choice of boundary condition
in $\Omega=(0,L)$ when $\Omega_1=(L/2-l/2, L/2+l/2)$ is chosen to have
half of the size of the interval. This result can be easily understood as a consequence of the  
fact that the entanglement entropy is basically due to the  behaviour of  field correlations 
at the interface between $\Omega_1$ and its complement 
$\Omega \hbox{\, $\backslash$ \,}  \Omega_1$ which does not involve the boundary values of the fields.
On the other hand  the  finite part $C_0$ is highly dependent  on the ultraviolet regularization
method.

 However, when  $\Omega_1$   reaches the boundary of the whole space $\Omega$ the entropy has the same
asymptotic behaviour \cite{calcar, calcar2}
\be
 S_l=\frac{C}{2} \log\frac{ l} { \epsilon} + \log g+ \frac12  C_0,
\label{eon}
\ee
but with a different coefficient  for the asymptotic logarithmic term 
and a different finite term  which is  related to the boundary entropy \cite{affleck} and, thus
also dependent on the boundary condition.
The behaviour of this quantity along the boundary renormalization group flow has  then
the same monotone behaviour that the boundary entropy.

A similar phenomenon occurs in 2+1 dimensions  with the constant term. In general, the
entropy is given by
\be
\quad  S_{\Omega_1}= C_1 \frac{L_1} {a}+ C \log \frac{L_1} {a}
+ C_0.
\ee
The logarithmic term is absent for domains $\Omega$ and $\Omega_1$
with  smooth boundaries $\partial\Omega$ and $\partial\Omega_1$,  whenever 
$\Omega\backslash\Omega_1$ is a connected manifold \cite{Fradkin}. In a regularized theory
the smoothness condition requires that the curvature of the boundaries  must be always much larger
than the ultraviolet cut-off $a$  \cite{hamma}. In that case, the remaining constant $C_0$  has a special behaviour 
because  not only is  regularisation independent but also  independent on the size of $\Omega_1$.  
$C_0$ can be split in two terms $C_0=C_0^\prime+C_0^\ast$, one $C_0^\prime$ which contains all
possible dependences on the  prescription used for the definition of the  $\Omega_1$ perimeter  $L_1$, and another
one $C_0^\ast$ which is absolutely prescription independent. In a massive theory,  if $L_1$ is much larger than the inverse 
of the mass gap $1/m$, there is a prescription which uniquely fixes the ambiguities involved in
such a splitting \cite{kitaev} \cite{wen}. If $\Omega_1$ is decomposed as the disjoint union of three similar domains
$\Omega_1=\Omega_\alpha\cup\Omega_\beta\cup \Omega_\gamma$, one can define
\be
C_0^\ast=C_0^{\Omega_1} -C_0^{\Omega_\alpha\cup\Omega_\beta}-C_0^{\Omega_\beta\cup\Omega_\gamma}
-C_0^{\Omega_\alpha\cup\Omega_\gamma}+C_0^{\Omega_\alpha}+C_0^{\Omega_\beta}+C_0^{\Omega_\gamma},
\ee
and the result is independent of the $\Omega_1$ decomposition and the perimeter definition prescription.
The constant $C_0^\ast$  is also shape independent and only really depends on the topology  of the domain 
$\Omega\, \backslash \, \Omega_1$. It   defines a topological invariant entropy $S_{\rm top}=C_0^\ast$ associated to the 
quantum vacuum \cite{kitaev} \cite{wen}, 
which measures its degree of  topological  entanglement. It can be shown that $S_{\rm top}=- \log \CD$, 
where $\CD$ is the total quantum dimension of the underlying topological theory. In our case  
case it is easy to show that  $\CD=1$, which means the vanishing of the topological entanglement entropy,
and that  result is independent of the boundary conditions. In more general
theories like the SU(2) WZWN theory  with level $k$ the topological entanglement
entropy is given by \cite{kitaev}
\be
 \displaystyle S_{\rm top}=\log \left[\sqrt{\frac{2}{k+2}}\sin\frac{\pi}{k+2}\right].
\ee
The quantum dimension $\CD$ is non-integer in that case   but
it is a  real topological invariant.

\section{Conclusions}

The novel thermodynamic quantities associated to field theories like boundary entropy and vacuum entanglement
entropy reveal new interesting properties of vacuum structure. The boundary entropy is associated
to the existence of boundary states and, thus, is very sensitive to the boundary conditions of the 
fields.  The role  of the vacuum entanglement entropy focuses on the measure of the amount 
of entanglement of the quantum vacuum and   is absolutely
independent of the type of boundary condition,  whenever  the domain where the
quantum fluctuations of the fields  are integrated out does not reach the boundary
of the space. However, when this domain  reaches the boundary, the entanglement entropy 
becomes dependent on the boundary conditions, displaying a monotone behaviour 
along the boundary renormalization group flow similar to that of the boundary entropy.

We have explicitly verified the   behaviour of boundary and entanglement entropies
under  changes of boundary conditions  for low dimensional massless free field theories.
The boundary entropy varies for quasiperiodic boundary conditions and Robin 
boundary conditions, whereas the entanglement entropy only changes
when the entanglement domain reaches the boundary or changes its topology. 
The same behaviour appears  in  three--dimensional field theories where the finite term
of the asymptotic behaviour of the entanglement entropy  can be related to a
new topological invariant (topological entanglement entropy). For free  scalar field
theories we have shown that this topological invariant is trivial for connected convex domains,
but  self-interacting field theories and non-connected domains 
might have  non-trivial topological entanglement entropy,  which  provides a basis for robust
codes in quantum computation \cite{kitaev}

In all  analysed cases  the  boundary entropy   does not increase along  the 
boundary  renormalization group flow from the 
ultraviolet to the infrared \cite{affleck}\cite{agm3}.
There are two interesting  problems which remain open: the effect of  interactions 
on both types of    entropies associated to the quantum  vacuum and their  behaviour 
for topological field theories. Both problems deserve  further analysis.

\section*{Acknowledgements}
We thank M. Aguado,  D. Garc\ii a Alvarez
and  J.I. Latorre, for interesting discussions. M.~A. also thanks the organizers
of the {\it 5th International Symposium on Quantum Theory and Symmetries}, and specially M. del Olmo,
for their hospitality in Valladolid during the meeting. This work is partially supported by CICYT (grant FPA2006-2315)
and DGIID-DGA (grant2007-E24/2).

\bigskip

\end{document}